\documentclass[a4paper,12pt]{article}

\usepackage{amsmath,amssymb,graphicx,amsthm,ifthen,epstopdf,fancyhdr}

\usepackage[bookmarks=true]{hyperref}
\usepackage{cite}
\usepackage{fancyhdr}
\usepackage{multirow}

\usepackage{array}
\usepackage[square,numbers,sort&compress]{natbib}
\usepackage{url}
\usepackage[usenames,dvipsnames]{xcolor}
\usepackage{color}
\hypersetup{
 colorlinks=true,
 citecolor=Blue,
 linkcolor=Red,
 urlcolor=Red
 }

\newcommand{\bX}{{\bf X}}

\newcommand{\cA}{{\cal A}}

\newcommand{\bitem}{\begin{itemize}}
\newcommand{\eitem}{\end{itemize}}
\newcommand{\goto}{\rightarrow}
\newcommand{\beq}{\begin{equation}}
\newcommand{\eeq}{\end{equation}}

\newcommand{\cM}{{\cal M}}
\newcommand{\mM}{{Mat}}
\newcommand{\mS}{{Sym}}
\def\Tr{{\rm Tr}}
\def\rank{{\rm rank}}
\def\cA{{\cal A}}
\def\reals{{\mathbb{R}}}
\def\normal{{\cal N}}
\newcommand{\mmx}{\mathcal{M}}

\title{The Phase Transition of Matrix Recovery from Gaussian
Measurements Matches the Minimax MSE of Matrix Denoising}

\author{
David L. Donoho \footnotemark[1]
\and Matan Gavish \footnotemark[1]
\and Andrea Montanari \footnotemark[1]\ \footnotemark[2]
}

\begin{document}
\maketitle

\renewcommand{\thefootnote}{\fnsymbol{footnote}}

\footnotetext[1]{Department of Statistics, Stanford University}
\footnotetext[2]{Department of Electrical Engineering, Stanford University}
\renewcommand{\thefootnote}{\arabic{footnote}}

\begin{abstract} 
Let $X_0$ be an unknown $M$ by $N$ matrix.  In matrix recovery,
  one takes $n < MN$ linear measurements $y_1, \dots , y_n$ of $X_0$, where $y_i
  = \Tr(a_i^T X_0)$ and each $a_i$ is a $M$ by $N$ matrix.  For measurement
  matrices with Gaussian i.i.d entries, it known that if $X_0$ is of low rank,
  it is recoverable from just a few measurements.  A popular approach for matrix
  recovery is Nuclear Norm Minimization (NNM): solving the convex optimization
  problem $\text{min }  \|X\|_* \text{ subject to } y_i=\Tr(a_i^TX)$ for all
  $1\le i\le n$, where $\|  \cdot \|_*$ denotes the nuclear norm, namely, the
  sum of singular values.  Empirical work reveals a \emph{phase transition}
  curve, stated in terms of the undersampling fraction $\delta(n,M,N) = n/(MN)$,
  rank fraction $\rho=r/N$ and aspect ratio $\beta=M/N$.  Specifically, a curve
  $\delta^* = \delta^*(\rho;\beta)$ exists such that, if $\delta >
  \delta^*(\rho;\beta)$, NNM typically succeeds, while if $\delta  <
  \delta^*(\rho;\beta)$, it typically fails.

  An apparently quite different problem is matrix denoising in Gaussian noise,
  where an unknown $M$ by $N$ matrix $X_0$ is to be estimated based on direct
  noisy measurements $Y =  X_0 + Z$, where the matrix $Z$ has iid Gaussian
  entries.  It has been empirically observed that, if $X_0$ has low rank, it may
  be recovered quite accurately from the noisy measurement $Y$.  A popular
  matrix denoising scheme solves the unconstrained optimization problem
  $\text{min }  \| Y - X \|_F^2/2 + \lambda \|X\|_* $.  When optimally tuned,
  this scheme achieves the asymptotic minimax MSE $\cM( \rho )  = \lim_{N \goto
  \infty} \inf_\lambda \sup_{\rank(X) \leq  \rho \cdot N}
  MSE(X,\hat{X}_\lambda)$.

  We report extensive experiments showing that the phase transition
  $\delta^*(\rho)$ in the first problem (Matrix Recovery from Gaussian
  Measurements) coincides with the minimax risk curve $\cM(\rho)$ in the second
  problem (Matrix Denoising in Gaussian Noise): $\delta^*(\rho) = \cM(\rho)$,
  for {\em any} rank fraction $0 < \rho < 1$. 

  Our experiments considered matrices belonging to two constraint classes: real
  $M$ by $N$ matrices, of various ranks and aspect ratios,  and real symmetric
  positive semidefinite $N$ by $N$ matrices, of various ranks.  Different
  predictions $\cM(\rho)$ of the phase transition location were used in the two
  different cases, and were validated by the experimental data.  
\end{abstract}

\tableofcontents

\section{Introduction}

Let $X_0$ be an unknown $M$ by $N$ matrix.  How many measurements must
we obtain in order to `completely know' $X_0$?  While it seems that $MN$
measurements must be necessary, in recent years intense research in applied
mathematics, optimization and information theory, has shown that, when $X_0$ is
of low rank, we may efficiently recover it from a relatively small number of
linear measurements by convex optimization
\cite{Recht2010,Candes2008,gross2011recovering}.  Applications have been
developed in fields ranging widely, for example from video and image processing
\cite{Candes2009}, to quantum state tomography \cite{Gross2010}, to
collaborative filtering \cite{Candes2008,keshavan2010matrix}.
 
Specifically, let $\cA:\reals^{M\times N}\to \reals^n$ be a linear operator and
consider measurements $y = \cA (X_0)$.  If $n < MN$, the problem of inferring
$X_0$ from $y$ may be viewed as attempting to solve an underdetermined system of
equations.  Under certain circumstances, it has been observed that this
(seemingly hopeless) task can be accomplished by solving the so-called nuclear
norm minimization problem \begin{equation} \label{NNMC} (P_{nuc}) \qquad
\text{min }  \|X\|_* \;\;\text{ subject to } y=\cA(X)\, .  \end{equation} Here
the \emph{nuclear norm} $\|X\|_*$  is the sum of singular values of $X$.  For
example, it was found that if $X_0$ is sufficiently low rank, with a principal
subspace  in a certain sense incoherent to the measurement operator $\cA$,
then the solution $X_1 = X_1(y)$ to $(P_{nuc})$ is precisely $X_0$. Such
incoherence can be obtained by letting $\cA$ be  random, for instance if
$\cA(X_0)_i = \Tr(a_i^{T}X_0)$ with $a_i\in\reals^{m\times n}$ having  i.i.d.
Gaussian entries.  In this case we speak of \emph{``matrix recovery from
Gaussian measurements''}  \cite{Recht2010}.

A key phrase from the previous paragraph: 
`if $X_0$ is \emph{sufficiently} low rank'.
Clearly there must be a quantitative trade-off between the
rank of $X_0$ and the number of measurements required,
such that higher rank matrices require more measurements.
In the Gaussian measurements model,
with $N$ sufficiently large, 
empirical work by Recht, Xu and Hassibi \cite{Recht2010a,Recht2008}, Fazel,
Parillo and Recht \cite{Recht2010},
  Tanner and Wei \cite{Tanner2012} and Oymak and Hassibi \cite{OymakHassibi11},
documents a  \emph{phase transition} phenomenon.
For matrices of a given rank, there is a fairly precise
number of required samples, in the sense that a transition 
from non recovery to complete recovery takes place sharply
as the number of samples varies across this value.
For example, in Figure \ref{table-01} below we report results obtained in
our own experiments, showing that, for reconstructing
matrices of size 60 by 60 which are of rank 20,  2600 Gaussian
measurements are sufficient with very high probability, but 
2400 Gaussian measurements are insufficient with very high probability.

\begin{figure}
  \begin{center}
    {\scriptsize
    \begin{tabular}{r| m{12pt}  m{12pt}  m{12pt}  m{12pt}  m{12pt}  m{12pt}  m{12pt}  m{12pt}  m{12pt}  m{12pt}  m{12pt}  m{12pt}  m{12pt} }
      \hline
      $\delta$ & 0.67 & 0.68 & 0.68 & 0.68 & 0.68 & 0.69 & 0.69 & 0.69 & 0.69 & 0.70 & 0.71 & 0.72 & 0.73 \\  
      $n$ & 2400 & 2437 & 2446 & 2455 & 2465 & 2474 & 2483 & 2492 & 2502 & 2511 & 2538 & 2575 & 2612 \\ 
      $\hat{\pi}(r|n,N)$ & 0.00 & 0.00 & 0.30 & 0.20 & 0.30 & 0.45 & 0.60 & 0.80 & 0.60 & 0.90 & 1.00 & 1.00 & 1.00 \\ 
      \hline
    \end{tabular}
    \caption{Data from typical Phase Transition experiment.
    Here $N=60$, $r=20$, and the number $n$ of Gaussian measurements varies.
    Note: our formula predicts an asymptotic phase transition at $\delta^* = 0.6937$,
    corresponding to $n=2497$.  And, indeed, the success probability is close to $1/2$ at that $n$.
    All runs involved $T=20$ Monte Carlo trials. }
    \label{table-01}
    }
  \end{center}
\end{figure}

In this paper, we present a simple and explicit formula for the phase
transition curve in matrix recovery from Gaussian measurements. The formula
arises in an apparently unrelated problem: matrix de-noising in Gaussian noise.
In this problem, we again let $X_0$ denote an $M$ by $N$ matrix, and we observe
$Y = X_0 + Z$, where $Z$ is Gaussian iid noise $Z_{ij}\sim\normal(0,1)$.  We
consider the following nuclear norm de-noising scheme:
\begin{align}
  (P_{nuc,\lambda}) \qquad  \min \Big\{\frac{1}{2} \| Y - X \|_F^2 +
  \lambda  \|X\|_* \Big\}\, .
\end{align}
In this problem the measurements $Y$ are direct, so in some sense complete, but
noisy.  The solution $\hat{X}_\lambda(Y)$ can be viewed as a shrinkage
estimator. In the basis of the singular vectors $U_Y$ and $V_Y$ of  $Y$, the
solution $\hat{X}_\lambda(Y)$ is diagonal, and the diagonal entries are produced
by soft thresholding of the singular values of $Y$. 

Because the measurements $y$ in the matrix recovery problem are noiseless but
incomplete, while the measurements $Y$ in the matrix denoising problem are
complete but noisy, the problems seem quite different. Nevertheless, we show
here that there is a deep connection between the two problems.  

\begin{figure}
  \centering
  \includegraphics[width=3in]{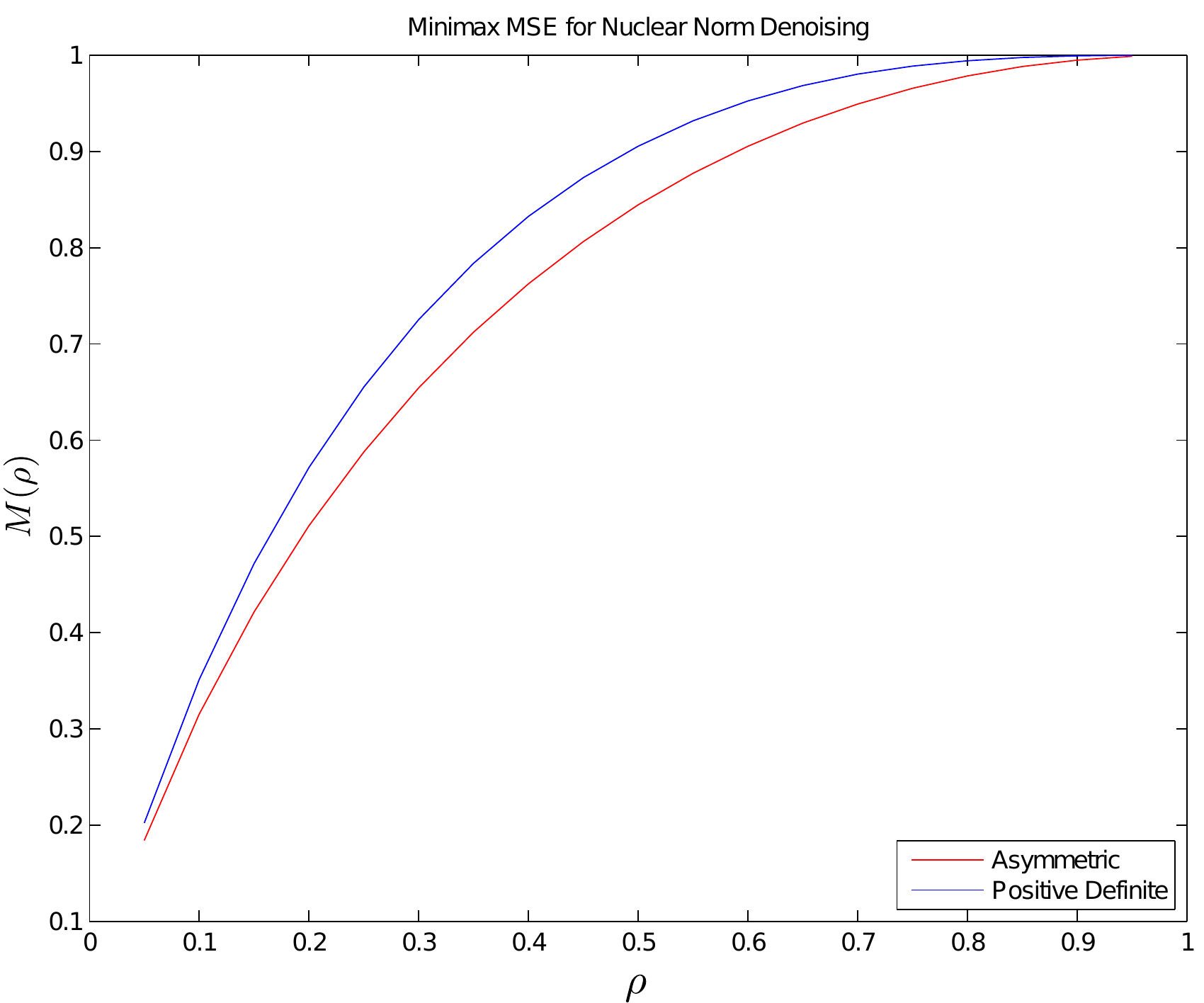}
  \caption{
  The two asymptotic minimax MSE's $M(\rho|\bX)$:
  $\bX = M_N$ (black), $\bX = Symm_N$ (red), in the case of square matrices.
  For non square matrices, see curves in Figure \ref{figure-DataCurves}}
  \label{figure-MMxMSE}
\end{figure}

\begin{figure}
  \centering
  \includegraphics[width=3in]{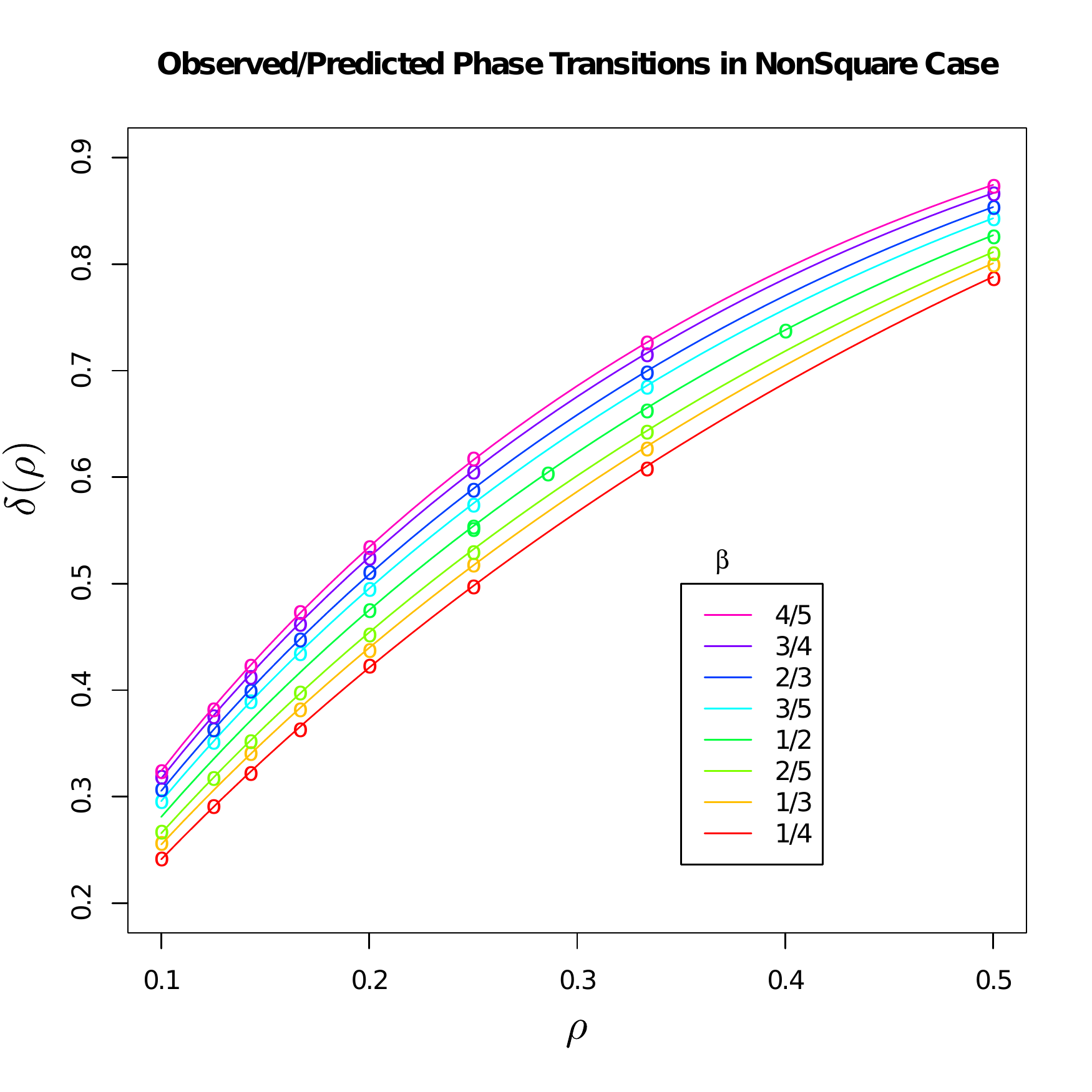}
  \caption{
  Curves:  Asymptotic Minimax MSE's  
  for nonsquare matrix settings $Mat_{M,N}$; 
  varying shape factor $\beta = M/N$.
  Points: Empirical phase transition locations for matrix recovery from
  incomplete measurements, see Table \ref{table-SI-NonSquare-Mat}.} 
  \label{figure-DataCurves}
\end{figure}

Let us quantify performance in the denoising problem by the minimax MSE, namely
\[
    \mmx(\rho;M,N) = \min_\lambda \max_{rank(X) \leq \rho N}
    MSE(X_0 \,,\,\hat{X}_\lambda(Y)), 
\]
where MSE refers to the dimension-normalized mean-squared error
\[
\frac{1}{MN}E \| X - \hat{X}_\lambda\|_F^2
\]
and  subscript $F$ denotes
Frobenius norm. The asymptotic minimax MSE 
$\mmx(\rho;\beta) = \lim_{N\goto \infty} \mmx(\rho;\beta N,N) $
has been derived in \cite{DonohoGavish2013}.
Explicit formulas for the curve $\rho\mapsto\mmx(\rho;\beta)$
appear in the Appendix. A parametric form is given for the case of asymptotically square matrices, 
$\beta=1$.
Figures \ref{figure-MMxMSE} and \ref{figure-DataCurves}  depict the
various minimax MSE curves.


\newpage
We can now state our main hypothesis for
 matrix recovery from Gaussian measurements.

{\bf Main Hypothesis: Asymptotic Phase Transition Formula.} \emph{Consider a sequence of matrix recovery problems 
with parameters $\{(r,n,M,N)\}_{N\ge 1}$ having limiting fractional
rank $\rho = \lim_{N\to\infty} r/N$,
limiting aspect ratio $\beta = \lim_{N\to\infty}M/N$,  and limiting
 incompleteness fraction $\delta = \lim_{N\to\infty} n/(MN)$. 
In the limit of large problem size $N$, the  solution $X_1(y)$ to the
nuclear norm minimization problem $(P_{nuc})$ 
 is correct with probability converging to one if $\delta > \cM(\rho;\beta)$ and
 incorrect with probability converging to one if $\delta < \cM(\rho;\beta)$.
}
In short: \emph{The asymptotic phase transition  $\delta^*(\rho,\beta)$ in
Gaussian matrix recovery
is equal to the asymptotic minimax MSE $\cM(\rho;\beta)$.}

In particular, for the case of small rank $r$, by studying the small
$\rho$ asymptotics of Eq. \eqref{argmin-func:eq}, we obtain that 
reconstruction is possible from  $n\ge 2r(M+N+\sqrt{MN})(1+O(r/N))$
measurements, but not from substantially less. 

This brief announcement tests this hypothesis by
conducting a substantial computer experiment generating large
numbers of random problem instances. We use statistical methods to check for
disagreement between the hypothesis and the predicted phase transition.  
To bolster the solidity of
our results, we conduct the experiment in two different settings:  $(1)$ the
matrix $X_0$ is a general $M$ by $N$ matrix, for various rank fractions $\rho$
and aspect ratios $\beta$; $(2)$ $X_0$ is a symmetric positive definite matrix,
for various rank fractions $\rho$.  In the latter case the positive semidefinite
constraint is added to the convex program $(P_{nuc})$.  As described below,
there are different asymptotic MSE curves for the two settings.  We demonstrate
an empirically accurate match in each of the cases, showing the depth and
significance of the connection we expose here. 

In the discussion and conclusion we connect our result with related work
in the field of sparsity-promoting reconstructions, where the same formal
identity between a minimax MSE and a phase transition boundary has
been observed, and in some cases even proved. We also discuss  recent
rigorous evidence towards establishing the above matrix recovery
phase transition formula.

\section{Methods}

We investigated the hypothesis
that the asymptotic phase transition boundary agrees with the proposed phase transition
formula to within experimental error.  

For notational simplicity we will focus here on the case $M=N$, and defer
the case of non-square matrices to the SI. Hence, we will drop
throughout the main text the argument $\beta=1$.
The asymptotic phase plane at point $(\rho,\delta)$
is associated to triples $(r,n,N)$, where $\rho  = r/N \in [0,1]$ is the rank fraction,
and $\delta = \delta(n,N|\bX) = n/dim(\bX)$ is the under sampling ratio, where $dim(\bX)$ is the dimension
of the underlying collection of matrices $\bX$.  We performed a sequence of experiments,
one for each tuple, in which we generated random rank-$r$ $N$ by $N$ matrices $X_0 \in \bX$,
random measurement matrices $\cA = A$ of size $n \times N^2$, and obtained random problem
instances $(y,A)$. We then applied a convex optimization procedure, obtaining a 
putative reconstruction $\hat{X} = \hat{X}(y,A)$. We declared a reconstruction successful
when the Frobenius norm was smaller than a threshold.
Our raw empirical observations  consist of a count of empirical successes  and sample sizes
at  a selected set of positions $(\rho,\delta)$ and a selected set of problem sizes $N$.
From these raw counts we produce fitted success probabilities $\hat{\pi}(r | n,N,\bX)$,
The finite-$N$ phase transition is the place where the true underlying 
probability of successful reconstruction take the value $50$\%. 
We tested the hypothesis that the finite-$N$
transition was consistent with the proposed asymptotic phase transition formula.

This section discusses details of data generation and data analysis.
 
\subsection{Generation of Problem Instances}

Each problem instance $(y,A)$ was generated by, first, generating  a random rank $r$ matrix
$X_0$, then, generating a random measurement matrix $A = A_{n,N^2}$ and then
applying $y = A \cdot  vec(X_0)$. 
 
 We considered problem instances of two specific types, corresponding to matrices $X_0 \in \bX$,
  with $\bX$ one of two classes of matrices
 \bitem
  \item $\mM_N$: all $N \times N$ matrices with real-valued entries
  \item $\mS_N$: all $N \times N$ real symmetric matrices which are 
    nonnegative-semidefinite
 \eitem

In the case $\bX = \mM_N$, we consider low-rank matrices 
$X_0 = U V'$ where $U$ and $V$ are each $N$ by $r$ partial orthogonal matrices in 
the Stiefel manifold $St(N,r)$.
The matrices are uniformly distributed on  $St(N,r)$. 
In the case $\bX = \mS_N$, we consider low-rank matrices 
$X_0 = U U'$ where $U$ is an $N$ by $r$ partial orthogonal matrix in $St(N,r)$,
and again the matrix is uniformly distributed. 

For measurement matrices $A$, we use 
Gaussian random matrices satisfying $A_{i,j} \sim N(0,1/n)$.

\subsection{Convex Optimization} 

For a given problem instance $(y,A)$, we attempt to recover the underlying low-rank
object $X_0$ from the measurements $y$ by  convex optimization.
Each of our choices $\bX$ gives rise to an associated optimization problem:
\[
  (P^\bX_{nuc}) \qquad  \text{min }  ||X||_* \text{ subject to } y=A \cdot vec(X), \qquad X \in \bX.
\]
 Here $\bX$ is one of these two classes of matrices $\mM_N$ or $\mS_N$.
The two resulting optimization problems can each 
be reduced to a so-called semidefinite programming problem;
see \cite{Fazel2001,Fazel2002}.

\subsection{Probability of Exact Recovery}

Since both the measurement matrix $A$, and 
the underlying low-rank object $X_0$
are random,  $(y,A)$ is a random
instance for $(P^\bX_{nuc})$.
The probability of exact recovery is defined by
\[
   \pi(r| n,N, \bX) = \text{Prob} \{  X_0 \mbox{ is the unique solution of }  (P^\bX_{nuc}) \}.
\]
Clearly $0 \leq \pi \leq 1$; for fixed $N$,
$\pi$ is monotone decreasing in $r$ and monotone increasing
in $n$. Also $\pi(r| n,N, \mM_N)  < \pi(r| n,N, \mS_N) $.

\subsection{Estimating the Probability of Exact Recovery}
 
Our procedure follows \cite{Donoho2009d,MoJaGaDo2013}.  
For a given  matrix type $\bX$ and rank $r$
 we conduct an experiment whose purpose is to estimate 
 $\pi(r|n,N,\bX)$ using $T$ Monte Carlo trials. 
In each trial we generate a random  instance 
$(y,A)$   which we supply to a solver for
$(P^\bX_{(nuc)})$,
obtaining the result $X_1$.  We compare the result $X_1$ to
$X_0$. If the relative error $\| X_0 - X_1 \|_F / \|X_0 \|_F$  is smaller than 
a numerical tolerance, we declare
the recovery a success;  if not, we declare it a failure. (In this paper, we
used an error tolerance of $0.001$.) We thus obtain $T$ binary measurements 
$Y_i$ indicating success or failure in reconstruction. The empirical
success fraction is then calculated as 
\[
\hat{\pi}(r | n,N, T, \bX) = \frac{\# \{ \text{successes} \} }{\# \{ \text{trials} \}}
= \frac{1}{T} \sum_{i=1}^T Y_i \,.
\]
These are the {\em raw observations} generated by our experiments. 

\subsection{Asymptotic Phase Transition Hypothesis}

Consider a sequence of tuples $(r,n,N)$ with $r/N \goto \rho$
 and $n/N \goto \delta$. We assume that there is 
 an asymptotic phase transition curve $\delta^*(\rho| \bX)$,
 i.e. a curve obeying 
\begin{equation} \label{asympPT}
    \pi(r|{n,N},\bX)  \goto \left \{  
            \begin{array}{ll} 
                  1 &  \delta < \delta^* ( \rho | \bX)  \\
                  0 &  \delta > \delta^* ( \rho | \bX) 
            \end{array} \right .
\end{equation}
For many convex optimization problems
the existence of such an asymptotic phase transition is rigorously proven;
see the Discussion below.
  
The hypothesis we investigate concerns the value of $\delta^*(\rho|\bX)$; specifically,
whether 
\begin{equation} \label{HypAsympPT}
    \delta^*(\rho|\bX) = \cM(\rho | \bX) .
\end{equation}
Here $\cM( \rho | \mM)$ (respectively $\cM( \rho | \mS)$ ) 
is the minimax MSE for SVT for general matrices (respectively, positive definite ones).
Formulas for $\cM$ were derived by the Authors
in \cite{DonohoGavish2013}; computational details are provided in the Appendix.
 
 \newcommand{\logit}{\mbox{logit}}
\subsection{Empirical Phase Transitions}
 The {\it empirical phase transition} point is estimated by fitting
a smooth function $\hat{\pi} (n/N)$ (in fact a logistic function) to the empirical data $\hat{\pi}(r | n,N,\bX)$
using the {\tt glm()} command in the R statistics language. In fact we fit the logistic model
that $\logit(\pi) \equiv \log(\frac{\pi}{1-\pi}) = a + b \Delta$, where $\Delta(\delta|\rho) = \delta - M(\rho)$ is the
offset between $\delta$ and the predicted phase transition.  The coefficients $a$ and $b$
are called the intercept and slope, and will be tabulated below. The intercept gives the
predicted logit exactly at $\Delta = 0$, i.e. $\delta = \cM(\rho)$.  
The empirical phase transition  is located at $\hat{\delta}(r,N,M,\bX)  = \cM(\rho) -a/b $. This is the value of $\delta = \delta(n,N|\bX)$ solving
\[
\hat{\pi} (\delta) =  1/2 .
\]
Under the hypothesis  (\ref{HypAsympPT}) 
we  have
\[
   \lim_{N\goto \infty, r/N \goto \rho} \lim_{T \goto \infty} \hat{\delta} (r,N,T,\bX)  = \cM(\rho | \bX).
\]
Consequently, in data analysis we will compare the fitted values  
$\hat{\delta}(r,N,T,\bX) $ with $\cM( r/N | \bX)$.

\subsection{Experimental Design}
To address our main hypothesis regarding the agreement of phase transition
boundaries, we measure $\hat{\pi}$
at points  $\delta = n/N$ and $\rho=r/N$
in the  phase plane $(\delta,\rho)$  which we expect to be maximally informative
about the location of the phase transition. In fact the informative
locations in binomial response models correspond to points where the probability of response
is in the middle range $(1/10,9/10)$ \cite{Kalish1990}. 
As a rough approximation to such an optimal design, we sample at equispaced  
$\delta \in [\cM(\rho|\bX) - 0.05, \cM(\rho|\bX) + 0.05]$.    

\subsection{Computing}
We primarily used  the MATLAB computing
environment, and the popular \textit{CVX} 
convex optimization package \cite{Grant2010}.
 A modeling system for disciplined convex programming by Boyd, Grant and others, supporting 
 two open source interior-point solvers: SeDuMi and SDPT3 
 \cite{Toh1999,Sturm1999}.

We also studied  the robustness of our results across solvers. 
Zulfikar Ahmed translated our code into Python and used the general purpose solver
package CVXOPT by Anderson and Vandeberghe \cite{Andersen2012}.

\section{Results}

Our experimental data have been deposited at \cite{pt-mmx-purl2013} they are
contained in a text file with more than 100,000 lines, each line reporting one
batch of Monte Carlo experiments at a given $r,n,M,N$ and $\bX$.  Each line also
documents the number of Monte Carlo trials $T$, and the observed success
fraction $\hat{\pi}$.  The file also contains metadata identifying the solver
and the researcher responsible for the run. 

 In all cases, we observed a transition from no observed successes
 at  $\delta = \cM(\rho) - 0.05$ to no observed failures at $\delta = \cM(\rho) + 0.05$.
 Figure \ref{figure-DataCurves} shows results we obtained at non square matrix ensembles,
 with varying $\beta=M/N$. The minimax MSE curves $M(\rho | \beta)$ vary widely,
 but the observed PT's track the curves closely.
 
\begin{figure}[h]
  \begin{center}
    \begin{tabular}{rrrrrrrr}
      \hline
      $\rho$ & $N$ & $T$ &$\cM(\rho)$ & $\hat{\delta}(\rho)$ & $a$ & $b$ & $Z$ \\ 
      \hline
      \multirow{5}{*}{1/10} &   40 &  400 & 0.351 & 0.352 & -0.128 & 182.978 & -0.581 \\ 
      &   50 &  400 & 0.351 & 0.350 & 0.282 & 200.131 & 1.213 \\ 
      &   60 &  400 & 0.351 & 0.352 & -0.096 & 221.212 & -0.398 \\ 
      &   80 &  400 & 0.351 & 0.350 & 0.415 & 295.049 & 1.452 \\ 
      &  100 &  400 & 0.351 & 0.349 & 0.641 & 383.493 & 1.900 \\
      \hline 
    \end{tabular}
    \caption{Results with $\bX=\mM_N$, with $\rho=1/10$ and Varying $N$. 
    $T$ total number of reconstructions; $a$, $b$: fitted logistic regression parameters;
    $Z$: traditional Z-score of logistic regression intercept.
    See Table \ref{table-SI-Mat}  for the complete table.
    }
    \label{table-Mat}
  \end{center}
\end{figure}

 Figure \ref{table-Mat} 
 shows a small subset of our results
 in the square case $\bX=\mM_N$, to explain our empirical results;
 the full tables are given in SI for the square, non square and
 symmetric positive definite cases.  
 In the square case, the empirical phase
 transition agrees in all cases with the formula $\cM(\rho)$ to two digits accuracy.
 Table \ref{table-SI-Sym} 
 shows that, in the symmetric 
 nonnegative-definite case $\bX=\mS_N$,  the empirical phase
 transition falls within  $[\cM(\rho)-0.01 , \cM(\rho)+0.01]$.
 
Previous empirical studies of phase transition behavior in sparse recovery
show that, even in cases where the asymptotic phase transition curve is rigorously proven
and analytically available, such large-$N$ theory
cannot be expected to match empirical finite-$N$ data to within the usual naive
standard errors \cite{Donoho2009d,MoJaGaDo2013}.
Instead, one observes  a finite transition zone of width $\approx c_1/N^{1/2}$ and
a small displacement of the empirical phase transition away from the asymptotic Gaussian phase transition, 
of size $\approx c_2/N$.  Hence, the strict literal device of testing the hypothesis that $E \hat{\delta} = \cM(\rho)$
is not appropriate in this setting \footnote{As shown in Figure \ref{table-Mat},
and in  Tables 
\ref{table-SI-Mat},\ref{table-SI-Sym},\ref{table-SI-NonSquare-Mat},\ref{table-SI-LowRank-Mat},and
\ref{table-SI-Rad},  
our results in many cases do generate
$Z$ scores for the intercept term in the logistic regression which are consistent with traditional
acceptance of this hypothesis. However,  traditional acceptance is not needed, in order for the
main hypothesis to be valid, and because of finite-$N$ scaling effects indicated above, would not ordinarily be expected to hold.}

A  precise statement of our hypothesis uses the fitted logistic parameters $\hat{a} = \hat{a}(r,N,M,\bX)$ 
and $\hat{b} =  \hat{b}(r,N,T,\bX)$, defined above.  The asymptotic relation\footnote{ $=_P$ denotes convergence in probability. }
\begin{equation} \label{EmpAsympPT}
        \lim_{N \goto \infty} \lim_{T \goto \infty}  \frac{\hat{a}(r,N,T,\bX)}{\hat{b}(r,N,T,\bX)} =_P 0, \qquad r = \lfloor \rho N \rfloor,
\end{equation}
implies $\delta(\rho|\bX) = \cM(\rho|\bX)$ .
Now note that in Figure \ref{table-Mat}  the coefficient $b$ scales
directly with $N$ and takes the value several hundred for large $N$.
This means that, in these experiments,
 the transition from complete failure to complete success
happens in a zone of width $< 1/100$. Notice also that
 $a$ stays bounded, for the most part between $0$ and $2$. This means that
 the response probability {\it evaluated exactly at $\cM(\rho)$} obeys typically:
 \[
        \logit(\hat{\pi}(\cM(\rho)) \in [0,2],
 \]

\begin{figure}[h]
  \centering
  \includegraphics[width=2.5in]{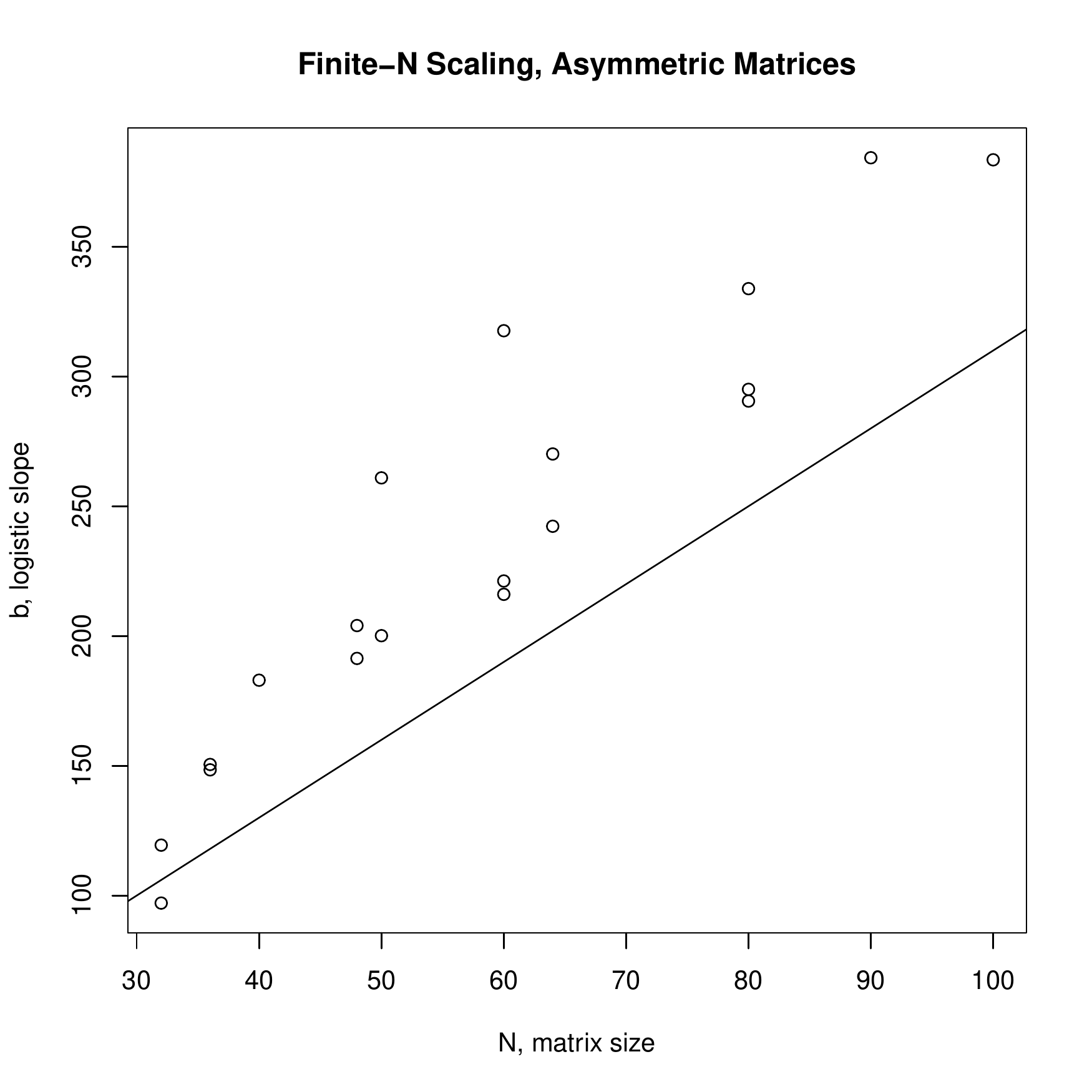} \qquad
  \includegraphics[width=2.5in]{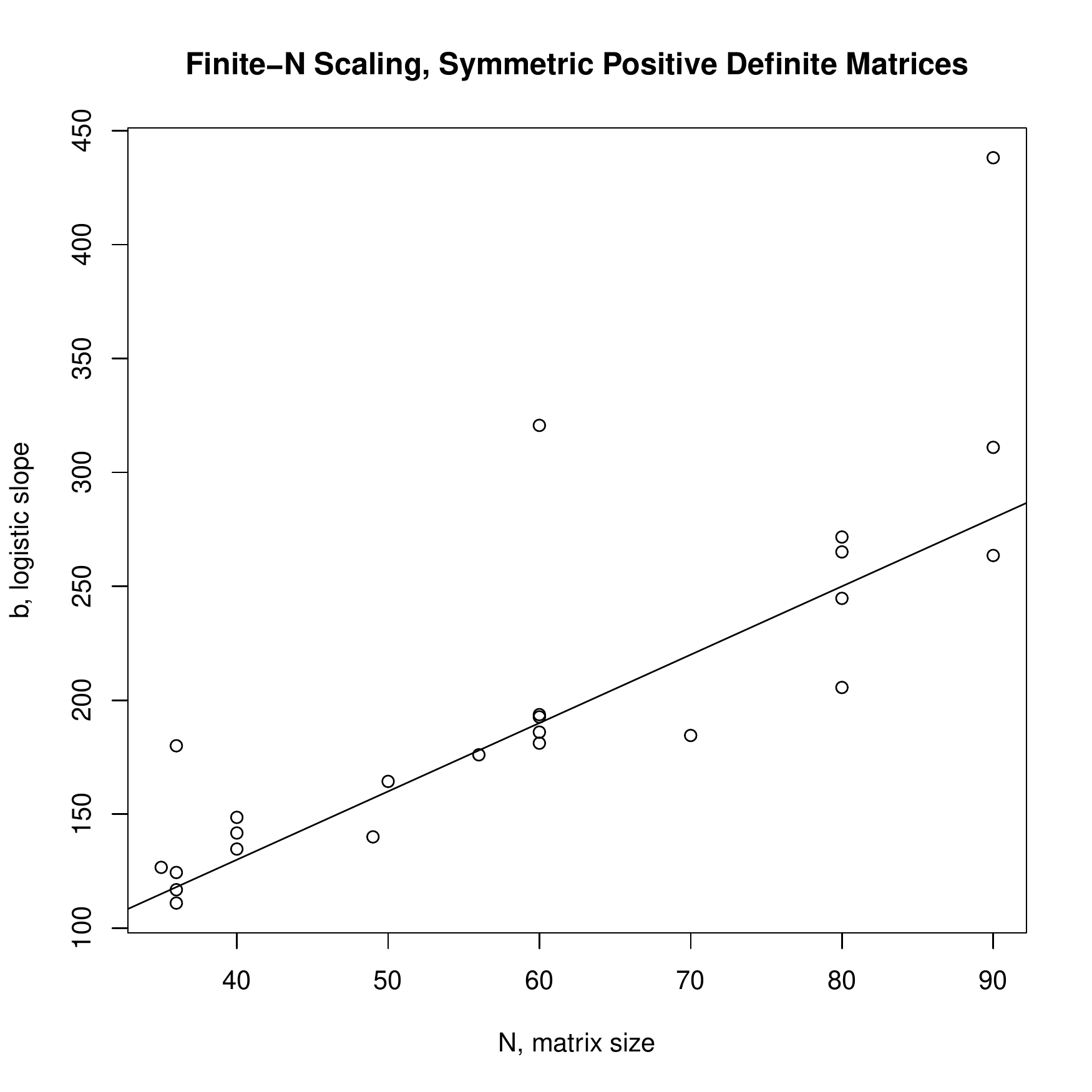} \qquad

  \caption{
  Finite-$N$ scaling effects.  Fitted slopes $b$ from Tables \ref{table-SI-Mat} and  
  \ref{table-SI-Sym}
  versus matrix sizes $N$.  
  Left Panel: asymmetric square matrices $\bX = \mM_N$.
  Right Panel: symmetric nonnegative-definite matrices $\bX = \mS_N$.
  Superimposed lines have formulas $b = 10 + 3 N$. Note:
  in Left Panel, data for $\rho \in \{ 3/4,9/10\}$ were excluded because the slopes
  $b$ were very large (700 and 2700, respectively).}
  \label{figure-N-scaling}
\end{figure}

Figure \ref{figure-N-scaling}, Panel A presents the fitted slopes $b$
and problem sizes $N$,as well as an empirical fit. All the data
came from Table \ref{table-SI-Mat}, 
but we omitted results for $\rho \in \{ 3/4,9/10\}$
because those slopes were large multiples of all other slopes.
Similarly Figure  \ref{figure-N-scaling}, Panel B presents the slopes
from Table \ref{table-SI-Sym}.  
In each plot, the line $b = 6 + 3 N$ is overlaid for comparison.
Both plots show a  clear tendency of the slopes to increase with $N$.

The combination of linear growth of  $b$ with $N$ and non-growth
of $a$  \footnote{Actually, even sub linear growth of $a$ implies the result.} 
implies Eq. \eqref{EmpAsympPT}, and  acceptance of our main hypothesis.

\paragraph{Nongaussian Measurements.} This paper studies matrix recovery from
Gaussian measurements of the unknown matrix, and specifically does not study
recovery from partial entry wise measurements typically called 'matrix completion'.
Entrywise measurements is yield a phase transition at a different
location \cite{Tanner2012}.  Our conclusions {\it do}
extend to certain nonGaussian measurements based on $\cA$ 
with independent and identically distributed entries
that are equiprobable $\pm 1$ 
(a.k.a. Rademacher matrices). See Table \ref{table-SI-Rad}.
 
\section{Discussion: Existing Literature and Our Contributions}

Phase transitions in the success of convex optimization
at solving non-convex problems have been observed previously.
Donoho and Tanner considered linear programs 
useful in the recovery of sparse vectors, 
rigorously established the existence of asymptotic phase transitions,
and derived exact formulas for  
the asymptotic phase transition 
\cite{Donoho2005,Do06,Donoho2009d,DoTa08,DoTa10}
as well as finite $N$ exponential bounds.
Their work considered the recovery of $k$-sparse
vectors with $N$ entries from $n$ Gaussian measurements.
The phase diagram in those papers
could be stated in terms
of variables $(\kappa,\delta)$, where, as in this paper, $\delta = n/N$ is
the under sampling fraction, and  $\kappa = k/N$ is the sparsity fraction,
which plays a role analogous to the role played here by the 
rank fraction $\rho$. The proofs in those papers were
obtained by techniques from high-dimensional combinatorial geometry,
and the formulas for the asymptotic phase transition were implicit,
written in terms of special functions.

Donoho, Maleki, and Montanari \cite{Donoho2009c} later developed
a new so-called Approximate Message Passing (AMP) 
approach to the
 sparse recovery problem, which gave new formulas for phase
 boundaries, confirmed rigorously by Bayati and Montanari
 in \cite{BM-MPCS-2011,BayatiMontanariLASSO}.
While the previous formulas involved combinatorial geometry, the new
(necessarily equivalent) ones involved instead minimax decision theory.
An extensive literature on AMP algorithms has since developed
see, e.g. \cite{SchniterEM,RanganGAMP}, implying, among
other things, universality in the sparse recovery phase transition \cite{Bayati2012Universality}.

Donoho, Johnstone and Montanari \cite{Donoho2011} generalized previous
AMP-based phase transition results to block-sparse, monotone, and
piecewise constant signals.
They provided evidence that 
the observed phase transition $\delta^*(\,\cdot\,)$ of the associated
convex optimization reconstruction methods
obeyed  formulas of the form
\begin{equation}
  \label{PT-Formula-Sparsity}
    \delta^*(\kappa) =\cM(\kappa),  \;\; 0 < \kappa < 1.
\end{equation}
Here $\kappa$ is a variable measuring generalized sparsity and $\cM(\kappa)$ 
the minimax MSE of an appropriate denoiser based on direct noisy measurements.
The main result in this brief report fits in this general framework whereby
the sparsity $\kappa$ is identified with the fractional rank $\rho$, and 
the minimax MSE symbol $\cM$ 
applies to  the singular value thresholding denoiser.
Our main result is therefore
an extension of  DJM-style formulas from the sparse recovery setting
problem to the low rank recovery setting.  

Much mathematical study of matrix recovery 
\cite{Candes2008,Recht2010a,Recht2010,OymakFazel2011}
has focused on providing rigorous bounds which show the existence
of a region of success, without however establishing a phase
transition phenomenon, or determining its exact boundary. 
A relatively accurate result  by Cand\`es and Recht (CR) for the case of Gaussian measurements
 \cite{candes2011simple},  implies that 
$n\ge 3r(M+N) [1+O(r/N)]$ measurements are sufficient. 
Our formulas for the square case $M=N$ show that
\[
    \cM(\rho | Mat_N) \sim 6 \rho,  \qquad \rho \goto 0.
\]
This agrees with the CR bound 
in the very low-rank case.  However, our relation
$\delta^*(\rho) = \cM(\rho)$ is apparently noticeably more
accurate than the CR formula at finite $N$.
Table \ref{table-SI-LowRank-Mat} 
presents experiments
where the rank is fixed at $r=1$,$2$,$3$, or $4$,
and $N$ varies between $40$ and $90$.  Even though in such cases
the corresponding $\rho=r/N$ is rather small, for example $1/90$ in the case
$r=1$ and $N=90$, the empirical PT  in our experiments agrees much more closely
with the nonlinear formula $\cM(\rho)$ than it does with the
linear formula $6 \rho$. Also, in the non square case $\beta \neq 1$,
$\cM \sim 2 \rho ( 1+ \beta + \sqrt{\beta})$, which is strictly
smaller than $6 \rho$ for $\beta < 1$, so the CR formula
is noticeably less accurate than the $ \delta^* = \cM$  formula in the non square case.

Of the  mathematical methods developed
to identify phase transitions but which are not based on
combinatorial geometry or approximate message passing,
the most precise are based on 
 the `Escape Through the Mesh' (ETM) technique of Yoram Gordon
\cite{Stojnic10a,OymakHassibi11}. ETM was used 
to prove upper bounds on the number of Gaussian measurements needed
for reconstruction of sparse signals  by Stojnic \cite{Stojnic10a} and for low-rank
matices by Oymak and Hassibi \cite{OymakHassibi11}.
In particular,  \cite{OymakHassibi11} studies one of the two
cases studied here and  observes in passing that  in the square case, ETM 
gives bounds that seemingly agree with actual phase transition
measurements.
Very recently, building on the same approach, a
DJM-style inequality $\delta^*(\kappa) \leq \cM(\kappa)$ has been announced
by Oymak and Hassibi for a wide range
of convex optimization problems in \cite{OymakHassibi12}, including
nuclear norm minimization; \cite{OymakHassibi12} also presented empirical evidence
for $\delta^*(\rho) = \cM(\rho)$ in the square case $\bX=\mM_N$.

{\it Our Contributions.} This paper presents explicit
formulas for the minimax MSE of singular value thresholding in various cases, and 
shows that the new formulas match the appropriate empirical phase transitions in
a formal comparison.  Compared to earlier work, we make here the following
contributions:

\bitem
\item {\it A Broad Range of Phase Transition Studies} for non square, square, and 
         symmetric nonnegative-definite matrix recovery from Gaussian measurements.
         We also made certain nonGaussian measurements and observed similar results.
  \item {\it A Broad Range of Prediction formulas.} We make available
explicit  formulas for Minimax MSE in the square symmetric
nonnegative-definite, or asymmetric case, as well as
non square cases.

 \item {\it Careful Empirical Technique},  including the following:
   \bitem 
      \item [] {\it Reproducibility.} Publication of the code and data underlying our conclusions.
      \item [] {\it Validation.}   Matlab/CVX results were re-implemented in Python / CVXOPT, with similar results.
                Code was executed on 3 different computing clusters, with similar results.
      \item [] {\it Study of Finite-$N$-scaling.} We studied tendencies  of $a$,$b$ as $N$ varies, and 
               observed behavior consistent with
                our asymptotic phase transition hypothesis.  Studies at a single $N$ could
                only have shown that an empirical phase transition was near to a given theoretical curve,
                at a given $N$, but not shown the scaling behavior with $N$ that the main hypothesis properly involves.
    \eitem
  \eitem

\section{Conclusions}

For the problem of matrix recovery from Gaussian measurements,
our experiments, as well as those of others, document the existence of a 
finite-$N$ phase transition. 
We compared our measured empirical phase transition curve with a formula
from the theory of matrix denoising and observed a compelling
match. Although the matrix recovery and matrix denoising problems
are superficially different, this match  evidences a deeper
connection, such that mean squared error properties of
a denoiser in a noise removal problem give precisely the exact recovery
properties of a  matrix recovery rule in a noiseless, but incomplete data problem.

This connection suggests both new limits on what is
possible in the matrix recovery problem, but also new
ways of trying to reach those limits.

\section*{Acknowledgments}
Thanks to Iain Johnstone for advice at several crucial points.
This work was partially supported by NSF DMS 0906812 (ARRA). MG was supported by 
a William R. and Sara Hart Kimball
Stanford Graduate Fellowship and a Technion EE Sohnis Promising Scientist Award.
AM was partially supported by the NSF CAREER award CCF-0743978 and 
the grant AFOSR/DARPA FA9550-12-1-0411.

\bibliographystyle{unsrtnat}

\appendix
\section{Asymptotic Minimax MSE Formula}

The following provides explicit formulas for the matrix denoising 
minimax curves $\mmx(\rho,\beta|Mat)$ and $\mmx(\rho|Sym)$ used above.
Please see \cite{DonohoGavish2013} for the derivations. Computer programs  
that efficiently calculate these quantities are provided in \cite{pt-mmx-purl2013}.
Let 
\begin{eqnarray} \label{MP:eq}
  P_{\gamma}(x;k)= \frac{1}{2\pi\gamma}\intop_x^{\gamma_+}
  t^{k-1}\sqrt{(\gamma_{+} - t)(t-\gamma_{-}}) \, dt\,,
\end{eqnarray}
where $\gamma_{\pm}=\left( 1\pm\sqrt{\gamma} \right)^2$,
 denote the complementary incomplete 
 moments of the Mar\u{c}enko-Pastur distribution \cite{Marcenko1967}.  
 Define
  \begin{eqnarray} \label{proxy:eq}
    & \mathbf{M}_\alpha(\Lambda;\rho,\tilde{\rho}) = 
    \rho + \tilde{\rho}  -  \rho\tilde{\rho}
    +   (1-\tilde{\rho}) 
    \Big[
    \rho \Lambda^2 + & \\
    & 
    + \alpha(1-\rho) \Big(
    P_\gamma( \Lambda^2 ; 1) -
    2\Lambda P_\gamma( \Lambda^2 ; \tfrac{1}{2}) 
    +  \Lambda^2  P_\gamma( \Lambda^2 ; 0)
    \Big) \Big]\,, & \nonumber
  \end{eqnarray}
  with $\gamma=\gamma(\rho,\tilde{\rho}) = (\tilde{\rho} - \rho\tilde{\rho}) / 
  (\rho-\rho\tilde{\rho})$.

  \paragraph{Case $\bX=\mM_{M,N}$.} 
  The minimax curve is given by
 $
 \mmx(\rho,\beta| \mM) = \inf_\Lambda \mathbf{M}_{1}(\Lambda; \rho,\beta\rho) .
 $
 The following minimaxity interpretation is proved in \cite{DonohoGavish2013}
\begin{eqnarray}  \label{nnm-pointwise-mmx:eq}
  \lim_{N\to\infty} \inf_{\lambda} \sup_{
  \stackrel{
  X_0 \in \reals^{\lfloor\beta N\rfloor \times N}}{rank(X_0) \leq \rho \beta N}
  }
 MSE(\hat{X}_\lambda,X_0) =
 \mmx(\rho,\beta | \mM)\,. 
\end{eqnarray}

\paragraph{Case $\bX=\mS_N$.} 
 The minimax curve is given by
 $
 \mmx(\beta| \mS) = \inf_\Lambda \mathbf{M}_{1/2}(\Lambda; \rho,\rho) .
 $
 The following minimaxity interpretation is proved in \cite{DonohoGavish2013}
\begin{eqnarray}  
  \lim_{N\to\infty} \inf_{\lambda} \sup_{
  \stackrel{
  X_0 \in S_N}{rank(X_0) \leq \rho N}
  }
 MSE(\hat{X}_\lambda,X_0) =
 \mmx(\rho | \mS)\,. 
\end{eqnarray}

\paragraph{Computing $\mmx(\rho,\beta | \bX)$.} 
The map $\Lambda\mapsto \mathbf{M}_\alpha(\Lambda;\rho,\tilde{\rho})$ 
is convex. Solving $d\mathbf{M}_\alpha / d\Lambda=0$ we get that
$
  \text{argmin}_\Lambda \, \mathbf{M}_\alpha(\Lambda;\rho,\tilde{\rho})
$
is the unique root of the equation
\begin{eqnarray} \label{argmin_Lambda:eq}
  \Lambda^{-1} P_\gamma(\Lambda^2;\tfrac{1}{2}) 
  - P_\gamma(\Lambda^2;0) = \frac{\rho}{\alpha(1-\rho)}\,. 
\end{eqnarray}
The right hand side of \eqref{argmin_Lambda:eq} is decreasing in $\Lambda$ and
the solution is determined numerically by binary search. 

For square matrices ($\rho=\tilde{\rho}$) \eqref{proxy:eq} can be expressed
using elementary trigonometric functions. In \cite{DonohoGavish2013} it is shown that
\begin{eqnarray} \label{proxy_sq:eq}
  &\mathbf{M}_\alpha(\Lambda;\rho,\rho) = 
    \rho(2-\rho)
    +   (1-\rho) \big[ \rho \Lambda^2 + & \\
    &
    + \alpha(1-\rho)\left(
    Q_2\left( \Lambda \right) -
    2\lambda Q_1\left( \Lambda\right) 
    +  \Lambda^2  Q_0\left( \Lambda\right)
    \right) \big]\,. &\nonumber
  \end{eqnarray}
where
\begin{scriptsize}
\begin{eqnarray*}
  \label{Q0:eq}
  Q_{0}(x) &=&
  \frac{1}{\pi} \intop_x^2 \sqrt{4-x^2} = 
   1 - \frac{x}{2\pi}\sqrt{4-x^2} -\frac{2}{\pi} atan(\frac{x}{\sqrt{4-x^2}} ) \\
 Q_{1}(x) &=&
  \frac{1}{\pi}\intop_x^2 x\sqrt{4-x^2} = 
    \frac{1}{3\pi}(4 - x^2)^{3/2} \\
    \label{Q2:eq}
  Q_{2}(x) &=& \frac{1}{\pi}\intop_x^2 x^2 \sqrt{4-x^2} =
  1 - \frac{1}{4\pi}x\sqrt{4-x^2}(x^2-2) - \frac{2}{\pi}asin(\frac{x}{2})
\end{eqnarray*}
\end{scriptsize}
are the complementary incomplete moments of the Quarter Circle law.
Moreover
\begin{eqnarray}
  \text{argmin}_\Lambda \mathbf{M}_\alpha(\Lambda;\rho,\rho) =
2 \cdot 
\sin\left( \theta_\alpha(\rho)\right)\,,
\end{eqnarray}
where $\theta_\alpha(\rho)\in[0,\pi/2]$ is the unique solution to the transcendental equation 
\begin{eqnarray} \label{argmin-func:eq}
  \theta + cot(\theta)\cdot\left( 1-\frac{1}{3}cos^2(\theta) \right)=
\frac{\pi(1+\alpha^{-1}\rho-\rho)}{2(1-\rho)}\,,
\end{eqnarray}
which is a simplified version of \eqref{argmin_Lambda:eq}.

\paragraph{Parametric representation of the minimax curves.}

For square matrices ($\rho=\tilde{\rho}$) the minimax curves
$\mmx(\rho,1|\mM)$ and $\mmx(\rho|\mS)$ admit a parametric representation in the
$(\rho,\mmx)$ plane using elementary trigonometric functions, see 
\cite{DonohoGavish2013}. As $\theta$ ranges over 
$[0,\pi/2]$, 
\begin{scriptsize}
\begin{eqnarray*}
  \rho(\theta) &=&  1 - \frac{\pi/2}{
  \theta+(\cot(\theta)\cdot(1-\frac{1}{3}cos^2(\theta)))} \\
\mmx(\theta) &=& 2\rho(\theta) - \rho^2(\theta) + 
4\rho(\theta)(1-\rho(\theta))sin^2(\theta) \\
&+& \frac{4}{\pi}(1-\rho)^2
\left[ 
(\pi-2\theta)(\frac{5}{4} - cos(\theta)^2) + \frac{sin(2\theta)}{12}(cos(2\theta)-14)
\right]
\end{eqnarray*}
\end{scriptsize}
is a parametric representation of $\mmx(\rho,1|\mM)$, and similarly
\begin{scriptsize}
\begin{eqnarray*}
  \rho(\theta) &=&  1 - 
  \frac{
 \theta+(\cot(\theta)\cdot(1-\frac{1}{3}cos^2(\theta))) - \pi/2
  }{
 \theta+(\cot(\theta)\cdot(1-\frac{1}{3}cos^2(\theta))) + \pi/2
  }\\
\mmx(\theta) &=& 2\rho(\theta) - \rho^2(\theta) + 
4\rho(\theta)(1-\rho(\theta))sin^2(\theta) \\
&+& \frac{2}{\pi}(1-\rho)^2
\left[ 
(\pi-2\theta)(\frac{5}{4} - cos(\theta)^2) + \frac{sin(2\theta)}{12}(cos(2\theta)-14)
\right]
\end{eqnarray*}
\end{scriptsize}
is a parametric representation of $\mmx(\rho|\mS)$.

\clearpage

\section{Summary of Empirical Results}

\begin{figure*}[h!]
\begin{center}
\begin{tabular}{rrrrrrrr}
  \hline
  $\rho$ & N & T &$\cM(\rho)$ & $\hat{\delta}(\rho)$ & $a$ & $b$ & $Z$ \\ 
  \hline
\multirow{5}{*}{1/10} &   40 &  400 & 0.351 & 0.352 & -0.128 & 182.978 & -0.581 \\ 
        &   50 &  400 & 0.351 & 0.350 & 0.282 & 200.131 & 1.213 \\ 
        &   60 &  400 & 0.351 & 0.352 & -0.096 & 221.212 & -0.398 \\ 
        &   80 &  400 & 0.351 & 0.350 & 0.415 & 295.049 & 1.452 \\ 
        &  100 &  400 & 0.351 & 0.349 & 0.641 & 383.493 & 1.900 \\
\hline 
\multirow{4}{*}{1/8} &   32 &  400 & 0.414 & 0.412 & 0.262 & 119.424 & 1.457 \\ 
&   48 &  400 & 0.414 & 0.413 & 0.262 & 204.025 & 1.120 \\ 
&   64 &  400 & 0.414 & 0.412 & 0.560 & 270.147 & 2.006 \\ 
&   80 &  400 & 0.414 & 0.413 & 0.296 & 290.536 & 1.057 \\ 
\hline
\multirow{2}{*}{1/6}  &   36 &  400 & 0.507 & 0.505 & 0.356 & 150.509 & 1.749 \\ 
&   60 &  400 & 0.507 & 0.506 & 0.137 & 216.067 & 0.571 \\ 
\hline
\multirow{4}{*}{1/4} &   32 &  400 & 0.655 & 0.651 & 0.398 & 97.107 & 2.375 \\ 
  &   48 &  400 & 0.655 & 0.653 & 0.356 & 191.373 & 1.554 \\ 
&   64 &  400 & 0.655 & 0.653 & 0.507 & 242.289 & 1.935 \\ 
&   80 &  400 & 0.655 & 0.651 & 1.312 & 333.842 & 3.584 \\ 
\hline
\multirow{4}{*}{1/3} 
 &   36 &  400 & 0.765 & 0.765 & -0.006 & 148.439 & -0.029 \\ 
 &   60 &  400 & 0.765 & 0.762 & 1.145 & 317.633 & 3.348 \\ 
 &   90 &  400 & 0.765 & 0.762 & 1.487 & 384.289 & 3.610 \\ 
\hline
  $1/2$ &   50 &  400 & 0.905 & 0.903 & 0.658 & 260.939 & 2.361 \\ 
 \hline
   $3/4$&   40 &  400 & 0.989 & 0.986 & 1.535 & 709.400 & 3.837 \\ 
 \hline
  $9/10$ &   50 &  400 & 0.999 & 0.998 & 2.478 & 2980.216 & 3.435 \\ 
   \hline
\end{tabular}
\caption{Results with $\bX=\mM_N$. }
\label{table-SI-Mat}
\end{center}
\end{figure*}

\begin{figure*}
\begin{center}
\begin{tabular}{rrrrrrrr}
  \hline
 $\rho$ & N & T & $\cM(\rho)$ & $\hat{\delta}(\rho)$ & $a$ & $b$ & $Z$ \\ 
  \hline
 \multirow{2}{*}{1/10}   &   40 &  800 & 0.315 & 0.310 & 0.787 & 148.605 & 4.268 \\ 
   &   60 &  800 & 0.315 & 0.312 & 0.519 & 186.020 & 2.640 \\ 
   &   80 &  800 & 0.315 & 0.311 & 0.995 & 265.114 & 3.864 \\ 
\hline 
\multirow{2}{*}{1/8} &   40 &  800 & 0.371 & 0.369 & 0.270 & 141.743 & 1.585 \\ 
    &   56 &  800 & 0.371 & 0.369 & 0.230 & 176.063 & 1.212 \\ 
    &   80 &  800 & 0.371 & 0.369 & 0.368 & 244.721 & 1.628 \\ 
\hline \multirow{2}{*}{1/7} &   35 &  800 & 0.407 & 0.403 & 0.623 & 126.659 & 3.730 \\ 
    &   49 &  800 & 0.407 & 0.404 & 0.496 & 140.034 & 2.878 \\ 
    &   70 &  800 & 0.407 & 0.404 & 0.581 & 184.556 & 2.898 \\ 
\hline\multirow{2}{*} {1/6} &   36 &  800 & 0.453 & 0.447 & 0.702 & 116.847 & 4.607 \\ 
    &   60 &  800 & 0.453 & 0.447 & 1.097 & 193.706 & 5.133 \\ 
    &   90 &  800 & 0.453 & 0.449 & 1.170 & 311.016 & 4.246 \\ 
\hline\multirow{2}{*}{1/5} &   40 &  800 & 0.511 & 0.505 & 0.786 & 134.698 & 5.448 \\ 
    &   50 &  800 & 0.511 & 0.507 & 0.707 & 164.398 & 4.528 \\ 
    &   80 &  800 & 0.511 & 0.507 & 0.859 & 205.621 & 4.778 \\ 
\hline\multirow{2}{*}{1/4}  &   36 &  800 & 0.588 & 0.579 & 0.967 & 110.989 & 7.120 \\ 
    &   60 &  800 & 0.588 & 0.582 & 1.155 & 181.174 & 6.379 \\ 
    &   80 &  800 & 0.588 & 0.583 & 1.450 & 271.651 & 6.066 \\ 
\hline\multirow{2}{*} {1/3} &   36 &  800 & 0.694 & 0.685 & 1.085 & 124.413 & 3.926 \\ 
    &   60 &  800 & 0.694 & 0.688 & 1.034 & 192.690 & 5.710 \\ 
    &   90 &  800 & 0.694 & 0.689 & 1.212 & 263.543 & 5.481 \\ 
\hline
\multirow{2}{*}{1/2} &   36 &  800 & 0.844 & 0.837 & 1.306 & 180.027 & 6.965 \\ 
    &   60 &  800 & 0.844 & 0.838 & 1.872 & 320.651 & 6.425 \\ 
    &   90 &  800 & 0.844 & 0.840 & 1.884 & 438.138 & 5.513 \\ 
   \hline
\end{tabular}
\caption{Results with $\bX=\mS_N$.}
\label{table-SI-Sym}
\end{center}
\end{figure*}

\clearpage

 \begin{figure*}
{\tiny
\begin{center}
\begin{tabular}{l|rrrrrrr}
  \hline
  $\beta$ & $\rho$ & $\cM(\rho)$& $\hat{\delta}(\rho)$& $a$ & $b$ & $Z$ & $\sqrt{M\cdot N}$ \\ 
  \hline
\multirow{7}{*}{$1/4$} 
 & 0.100 & 0.241 & 0.241 & 0.020 & 215.506 & 0.085 & 60.000 \\ 
 & 0.125 & 0.290 & 0.290 & -0.037 & 322.468 & -0.127 & 64.000 \\ 
 & 0.143 & 0.323 & 0.321 & 0.468 & 238.477 & 1.813 & 70.000 \\ 
 & 0.167 & 0.365 & 0.362 & 0.884 & 247.690 & 3.109 & 60.000 \\ 
 & 0.200 & 0.421 & 0.422 & -0.153 & 246.541 & -0.597 & 60.000 \\ 
 & 0.250 & 0.498 & 0.496 & 0.339 & 214.134 & 1.405 & 64.000 \\ 
 & 0.333 & 0.610 & 0.607 & 0.856 & 284.629 & 2.834 & 60.000 \\ 
 & 0.500 & 0.788 & 0.786 & 0.635 & 297.803 & 2.124 & 64.000 \\
\hline 
\multirow{7}{*}{$1/3$}
  & 0.100 & 0.255 & 0.255 & -0.105 & 269.362 & -0.395 & 51.962 \\ 
  & 0.143 & 0.340 & 0.340 & 0.031 & 210.274 & 0.133 & 48.497 \\ 
  & 0.167 & 0.383 & 0.381 & 0.373 & 169.227 & 1.733 & 51.962 \\ 
  & 0.200 & 0.440 & 0.437 & 0.589 & 197.335 & 2.457 & 51.962 \\ 
  & 0.250 & 0.517 & 0.517 & 0.113 & 218.479 & 0.470 & 55.426 \\ 
  & 0.333 & 0.629 & 0.626 & 0.661 & 219.930 & 2.584 & 51.962 \\ 
  & 0.500 & 0.801 & 0.799 & 0.572 & 274.122 & 2.036 & 55.426 \\ 
\hline
\multirow{7}{*}{$1/2$} 
 & 0.200 & 0.475 & 0.474 & 0.127 & 159.697 & 0.616 & 42.426 \\ 
 & 0.250 & 0.554 & 0.553 & 0.258 & 201.035 & 1.114 & 45.255 \\ 
 & 0.286 & 0.604 & 0.603 & 0.307 & 167.955 & 1.434 & 49.497 \\ 
 & 0.333 & 0.665 & 0.662 & 0.486 & 155.704 & 2.327 & 42.426 \\ 
 & 0.400 & 0.738 & 0.737 & 0.292 & 184.091 & 1.308 & 42.426 \\ 
 & 0.500 & 0.827 & 0.825 & 0.460 & 228.287 & 1.826 & 45.255 \\ 
 & 0.667 & 0.930 & 0.927 & 0.729 & 248.019 & 3.073 & 42.426 \\
\hline 
\multirow{8}{*}{$3/5$}
 & 0.100 & 0.296 & 0.295 & 0.146 & 191.950 & 0.646 & 38.730 \\ 
 & 0.125 & 0.352 & 0.351 & 0.242 & 219.767 & 1.000 & 61.968 \\ 
 & 0.143 & 0.389 & 0.388 & 0.105 & 179.344 & 0.484 & 54.222 \\ 
 & 0.167 & 0.436 & 0.433 & 0.425 & 198.788 & 1.814 & 46.476 \\ 
 & 0.200 & 0.495 & 0.494 & 0.237 & 146.409 & 1.196 & 38.730 \\ 
 & 0.250 & 0.575 & 0.573 & 0.361 & 164.339 & 1.704 & 46.476 \\ 
 & 0.333 & 0.686 & 0.684 & 0.509 & 258.045 & 1.886 & 46.476 \\ 
 & 0.500 & 0.843 & 0.842 & 0.153 & 156.137 & 0.751 & 46.476 \\
\hline 
\multirow{8}{*}{$2/3$}
& 0.100 & 0.305 & 0.306 & -0.168 & 221.047 & -0.697 & 36.742 \\ 
& 0.125 & 0.363 & 0.362 & 0.064 & 151.402 & 0.321 & 39.192 \\ 
& 0.143 & 0.401 & 0.399 & 0.263 & 142.052 & 1.346 & 34.293 \\ 
& 0.167 & 0.448 & 0.446 & 0.271 & 157.783 & 1.323 & 36.742 \\ 
& 0.200 & 0.509 & 0.510 & -0.209 & 155.201 & -1.034 & 36.742 \\ 
& 0.250 & 0.589 & 0.587 & 0.315 & 152.493 & 1.555 & 39.192 \\ 
& 0.333 & 0.700 & 0.697 & 0.408 & 190.613 & 1.768 & 36.742 \\ 
& 0.500 & 0.854 & 0.853 & 0.257 & 232.865 & 1.034 & 39.192 \\
\hline 
\multirow{8}{*}{$3/4$}
 & 0.100 & 0.317 & 0.317 & -0.043 & 163.597 & -0.207 & 34.641 \\ 
 & 0.125 & 0.376 & 0.375 & 0.237 & 192.933 & 1.047 & 55.426 \\ 
 & 0.143 & 0.415 & 0.412 & 0.562 & 178.363 & 2.476 & 48.497 \\ 
 & 0.167 & 0.463 & 0.461 & 0.304 & 174.599 & 1.405 & 41.569 \\ 
 & 0.200 & 0.525 & 0.524 & 0.159 & 120.459 & 0.886 & 34.641 \\ 
 & 0.250 & 0.606 & 0.604 & 0.276 & 154.259 & 1.353 & 41.569 \\ 
 & 0.333 & 0.716 & 0.714 & 0.378 & 180.963 & 1.682 & 41.569 \\ 
 & 0.500 & 0.867 & 0.866 & 0.341 & 283.911 & 1.227 & 41.569 \\
\hline 
\multirow{8}{*}{$4/5$}
 & 0.100 & 0.324 & 0.323 & 0.220 & 174.269 & 1.019 & 44.721 \\ 
 & 0.125 & 0.384 & 0.381 & 0.423 & 160.001 & 1.999 & 44.721 \\ 
 & 0.143 & 0.423 & 0.422 & 0.273 & 248.331 & 1.055 & 62.610 \\ 
 & 0.167 & 0.472 & 0.472 & -0.025 & 190.094 & -0.112 & 40.249 \\ 
 & 0.200 & 0.534 & 0.533 & 0.204 & 195.806 & 0.889 & 44.721 \\ 
 & 0.250 & 0.616 & 0.616 & -0.076 & 197.138 & -0.333 & 44.721 \\ 
 & 0.333 & 0.726 & 0.725 & 0.213 & 203.942 & 0.910 & 40.249 \\ 
 & 0.500 & 0.875 & 0.873 & 0.412 & 215.351 & 1.689 & 44.721 \\ 
   \hline
\end{tabular}
\caption{Results with non square matrices $\bX=\mM_{M,N}$.
Each row based on $T=400$ Monte Carlo trials. }
\label{table-SI-NonSquare-Mat}
\end{center}
}
\end{figure*}

\clearpage

 \begin{figure*}
\begin{center}
\begin{tabular}{rrrrrrrrr}
  \hline
  N & r & $\rho$ & $6\rho$ & $\cM(\rho)$ & $\hat{\delta}(\rho)$ & $a$ &$b$ & $Z$ \\ 
   \hline
   90 & 1 & 0.011 & 0.067 & 0.059 & 0.054 & 3.595 & 828.597 & 4.627 \\ 
   80 & 1 & 0.013 & 0.075 & 0.065 & 0.059 & 2.766 & 534.431 & 5.117 \\ 
   70 & 1 & 0.014 & 0.086 & 0.072 & 0.069 & 1.444 & 432.528 & 3.962 \\ 
   60 & 1 & 0.017 & 0.100 & 0.081 & 0.078 & 1.374 & 434.911 & 3.542 \\ 
   50 & 1 & 0.020 & 0.120 & 0.094 & 0.091 & 1.159 & 348.687 & 3.226 \\ 
 90 & 2 & 0.022 & 0.133 & 0.103 & 0.101 & 1.225 & 611.400 & 2.499 \\ 
 40 & 1 & 0.025 & 0.150 & 0.114 & 0.114 & -0.169 & 328.547 & -0.513 \\ 
 40 & 1 & 0.025 & 0.150 & 0.114 & 0.114 & -0.169 & 328.547 & -0.513 \\ 
 70 & 2 & 0.029 & 0.171 & 0.127 & 0.125 & 6.383 & 2746.893 & 0.007 \\ 
 30 & 1 & 0.033 & 0.200 & 0.145 & 0.145 & 0.013 & 137.638 & 0.051 \\ 
 30 & 1 & 0.033 & 0.200 & 0.145 & 0.145 & 0.013 & 137.638 & 0.051 \\ 
 30 & 1 & 0.033 & 0.200 & 0.145 & 0.145 & 0.013 & 137.638 & 0.051 \\ 
 80 & 3 & 0.037 & 0.225 & 0.160 & 0.158 & 0.493 & 345.100 & 1.111 \\ 
 50 & 2 & 0.040 & 0.240 & 0.169 & 0.166 & 1.071 & 364.190 & 1.981 \\ 
 70 & 3 & 0.043 & 0.257 & 0.179 & 0.177 & 3.442 & 1883.417 & 0.006 \\ 
 90 & 4 & 0.044 & 0.267 & 0.184 & 0.183 & 2.044 & 1773.906 & 0.006 \\ 
 20 & 1 & 0.050 & 0.300 & 0.203 & 0.203 & -0.055 & 94.130 & -0.219 \\ 
 20 & 1 & 0.050 & 0.300 & 0.203 & 0.203 & -0.055 & 94.130 & -0.219 \\ 
 20 & 1 & 0.050 & 0.300 & 0.203 & 0.203 & -0.055 & 94.130 & -0.219 \\ 
 20 & 1 & 0.050 & 0.300 & 0.203 & 0.203 & -0.055 & 94.130 & -0.219 \\ 
 70 & 4 & 0.057 & 0.343 & 0.226 & 0.227 & -2.755 & 2732.891 & -0.001 \\ 
 50 & 3 & 0.060 & 0.360 & 0.235 & 0.230 & 6.462 & 1302.371 & 0.008 \\ 
 30 & 2 & 0.067 & 0.400 & 0.256 & 0.255 & 0.106 & 152.849 & 0.290 \\ 
 30 & 2 & 0.067 & 0.400 & 0.256 & 0.255 & 0.106 & 152.849 & 0.290 \\ 
 40 & 3 & 0.075 & 0.450 & 0.281 & 0.280 & 0.114 & 191.765 & 0.249 \\ 
 50 & 4 & 0.080 & 0.480 & 0.296 & 0.294 & 0.361 & 225.965 & 0.622 \\ 
 20 & 2 & 0.100 & 0.600 & 0.351 & 0.345 & 0.387 & 60.051 & 1.345 \\ 
 20 & 2 & 0.100 & 0.600 & 0.351 & 0.345 & 0.387 & 60.051 & 1.345 \\ 
 20 & 2 & 0.100 & 0.600 & 0.351 & 0.345 & 0.387 & 60.051 & 1.345 \\ 
 30 & 4 & 0.133 & 0.800 & 0.434 & 0.434 & -0.005 & 115.737 & -0.011 \\ 
 20 & 3 & 0.150 & 0.900 & 0.472 & 0.480 & -0.690 & 86.275 & -1.510 \\ 
 20 & 4 & 0.200 & 1.200 & 0.572 & 0.587 & -0.994 & 64.706 & -2.057 \\ 
   \hline
\end{tabular}
\caption{Results with low rank square matrices $\bX=\mM_{N,N}$. 
Each row based on $T=400$ Monte Carlo trials.}
\label{table-SI-LowRank-Mat}
\end{center}
\end{figure*}

\begin{figure*}
\begin{center}
\begin{tabular}{rrrrrrr}
  \hline
  $\rho$ & $N$ & $\cM(\rho)$ & $\hat{\delta}(\rho)$ & $a$ & $b$ & $Z$ \\ 
  \hline
\multirow{3}{*}{$1/10$} & 40.000 & 0.351 & 0.350 & 0.187 & 170.554 & 0.878 \\ 
 & 60.000 & 0.351 & 0.350 & 0.239 & 286.765 & 0.863 \\ 
 & 80.000 & 0.351 & 0.351 & 0.166 & 290.560 & 0.597 \\ 
\hline
\multirow{3}{*}{$1/8$} & 40.000 & 0.414 & 0.413 & 0.145 & 146.920 & 0.736 \\ 
 & 56.000 & 0.414 & 0.412 & 0.552 & 215.088 & 2.220 \\ 
 & 80.000 & 0.414 & 0.413 & 0.667 & 375.205 & 1.992 \\ 
\hline
\multirow{3}{*}{$1/7$} & 35.000 & 0.456 & 0.457 & -0.158 & 141.742 & -0.813 \\ 
 & 49.000 & 0.456 & 0.455 & 0.158 & 216.040 & 0.659 \\ 
 & 70.000 & 0.456 & 0.454 & 0.911 & 394.564 & 2.527 \\ 
\hline
\multirow{3}{*}{$1/6$} & 36.000 & 0.507 & 0.505 & 0.370 & 159.357 & 1.768 \\ 
 & 60.000 & 0.507 & 0.505 & 0.570 & 245.893 & 2.137 \\ 
 & 90.000 & 0.507 & 0.505 & 0.689 & 335.299 & 2.168 \\ 
 \hline
\multirow{3}{*}{$1/5$} & 40.000 & 0.572 & 0.571 & 0.089 & 172.047 & 0.417 \\ 
 & 60.000 & 0.572 & 0.568 & 0.912 & 248.273 & 3.188 \\ 
 & 80.000 & 0.572 & 0.570 & 0.598 & 285.542 & 2.073 \\ 
\hline
\multirow{3}{*}{$1/3$} & 36.000 & 0.765 & 0.759 & 0.752 & 124.479 & 3.825 \\ 
 & 60.000 & 0.765 & 0.763 & 0.608 & 234.115 & 2.324 \\ 
 & 81.000 & 0.765 & 0.762 & 1.053 & 288.525 & 3.306 \\ 
\hline
\multirow{3}{*}{$1/2$}& 36.000 & 0.905 & 0.903 & 0.487 & 229.398 & 1.914 \\ 
 & 60.000 & 0.905 & 0.902 & 1.403 & 379.105 & 3.518 \\ 
 & 90.000 & 0.905 & 0.903 & 11.365 & 3971.012 & 0.008 \\ 
    \hline
\end{tabular}
\caption{Results with Rademacher measurements of square matrices $\bX=\mM_{N,N}$. 
Each row based on $T=400$ Monte Carlo trials.}
\label{table-SI-Rad}
\end{center}
\end{figure*}

\clearpage

\section{Data Deposition}

The data have been deposited in a text file at  \cite{pt-mmx-purl2013}.
A typical fragment of the file is given here:
{\tiny
\begin{verbatim}
Line Project             Experiment      M  N  S Instance rank  rho              delta            Err0       Err1  Err2
3381 Nuc_CVX_20121129dii Nuc_CVX_N12S01m 12 12 1 m 4 0.333333333333333 0.73395061728395 0.016630719182629 0 0.444444444444444
3382 Nuc_CVX_20121129dii Nuc_CVX_N12S01m 12 12 1 m 4 0.333333333333333 0.739213775178687 0.0159010475527301 0 0.465277777777778
3383 Nuc_CVX_20121129dii Nuc_CVX_N12S01m 12 12 1 m 4 0.333333333333333 0.744476933073424 0.0161172486497232 0 0.416666666666667
3384 Nuc_CVX_20121129dii Nuc_CVX_N12S01m 12 12 1 m 4 0.333333333333333 0.749740090968161 0.00149080158529591 1 1
3385 Nuc_CVX_20121129dii Nuc_CVX_N12S01m 12 12 1 m 4 0.333333333333333 0.755003248862898 0.0340839945130298 0 0.173611111111111
3386 Nuc_CVX_20121129dii Nuc_CVX_N12S01m 12 12 1 m 4 0.333333333333333 0.760266406757635 0.0235186056093925 0 0.361111111111111
3387 Nuc_CVX_20121129dii Nuc_CVX_N12S01m 12 12 1 m 4 0.333333333333333 0.765529564652372 0.0136400454215757 0 0.506944444444444
3388 Nuc_CVX_20121129dii Nuc_CVX_N12S01m 12 12 1 m 4 0.333333333333333 0.770792722547109 1.11644848368808e-09 1 1
3389 Nuc_CVX_20121129dii Nuc_CVX_N12S01m 12 12 1 m 4 0.333333333333333 0.776055880441845 0.00194118165102407 0 1
3390 Nuc_CVX_20121129dii Nuc_CVX_N12S01m 12 12 1 m 4 0.333333333333333 0.781319038336582 0.0059943065062774 0 0.902777777777778
3391 Nuc_CVX_20121129dii Nuc_CVX_N12S01m 12 12 1 m 4 0.333333333333333 0.786582196231319 1.83878516274726e-09 1 1
3392 Nuc_CVX_20121129dii Nuc_CVX_N12S01m 12 12 1 m 4 0.333333333333333 0.791845354126056 7.85225405546251e-09 1 1
3393 Nuc_CVX_20121129dii Nuc_CVX_N12S01m 12 12 1 m 4 0.333333333333333 0.797108512020793 4.47138355029567e-10 1 1
3394 Nuc_CVX_20121129dii Nuc_CVX_N12S01m 12 12 1 m 4 0.333333333333333 0.80237166991553 1.03566257175607e-08 1 1
3395 Nuc_CVX_20121129dii Nuc_CVX_N12S01m 12 12 1 m 4 0.333333333333333 0.807634827810266 2.62954112892325e-09 1 1
3396 Nuc_CVX_20121129dii Nuc_CVX_N12S01m 12 12 1 m 4 0.333333333333333 0.812897985705003 0.00120410713874671 1 1
3397 Nuc_CVX_20121129dii Nuc_CVX_N12S01m 12 12 1 m 4 0.333333333333333 0.81816114359974 1.03259685844506e-08 1 1
3398 Nuc_CVX_20121129dii Nuc_CVX_N12S01m 12 12 1 m 4 0.333333333333333 0.823424301494477 2.52202641519081e-10 1 1
3399 Nuc_CVX_20121129dii Nuc_CVX_N12S01m 12 12 1 m 4 0.333333333333333 0.828687459389214 1.94437637948599e-09 1 1
3400 Nuc_CVX_20121129dii Nuc_CVX_N12S01m 12 12 1 m 4 0.333333333333333 0.83395061728395 1.69829115472996e-10 1 1
\end{verbatim}
}

The fields have the following meaning
\bitem
 \item Line -- Line number in file; in the above example, lines 3381-3400.
 \item Project -- File identifier -- allows identification of code and logs that generated these data; in the above fragment, 
 {\tt 'Nuc\_CVX\_20121129dii'}.
 \item Experiment -- File identified -- allows identification of code and logs that generated these data; 
 in the above fragment {\tt 'Nuc\_CVX\_N12S01m'}.
 \item M,N -- matrix size of $X_0$, i.e $M$ by $N$ matrix; in the above fragment $M=N=12$.
 \item S -- number of matrices in a stack (see below)
 \item Instance -- alphabetic code a-t, identifying one of 20 identical runs which generated this result; in the above fragment, 'm'.
 \item rank -- integer rank of matrix; in the above fragment 'm'.
 \item rho -- fraction in $[0,1]$, $\rho = \mbox{rank}/N$; in the above fragment, $1/3$.
 \item delta -- $\delta = n/(MN)$ in case of asymmetric matrix, or $\delta=2*n/(N(N+1))$ in case of symmetric matrix.
 \item Err0 --  $\|\hat{X}-X_0\|_F/(NM)^{1/2}$.
 \item Err1 --  1 iff $\|\hat{X}-X_0\|_F/\|X_0\|_F < tol$, and $0$ otherwise
 \item Err2 --  fraction of entries with discrepancy $|\hat{X}(i,j)-X_0(i,j) | < tol$.
\eitem
Additional concepts:
\bitem
 \item {\it Numerical Tolerance.} In our experiments, we used a numerical error tolerance parameter $tol = 0.001$.  
 \item Our experiments also extensively covered cases where $X_0$ is a 'stack of matrices', i.e. a 3-way array $M \times N \times S$,
  where $S$ is the number of items in the stack.  The only cases of interest for this paper are $S=1$.
\eitem

\section{Code Deposition}

There are two types of code deposition.
\bitem
 \item {\it Reproduction from Data Deposition.} 
 The code that actually makes the figures and tables we presented in this paper,
 starting from the data deposition.  This is deposited at \cite{pt-mmx-purl2013}.
 The code we actually ran to create our figures and tables is a set of R
 scripts, and was run on a Mac OS X.  We believe the same code runs with
 minimal changes on a LINUX environment.
 \item {\it RunMyCode Deposition.}
 For readers who wish simply to compute {\it the value of the Minimax Mean-Squared Error}
 over each of the matrix classes we considered, we offer a Minimax MSE calculator at
 RunMyCode.org.
 \item {\it Full Code and Results Deposition.}
 At 
  \cite{pt-mmx-purl2013},
 we also offer a literal dump of the code we ran and
 all the logs and result files we obtained.
\eitem
We believe the first two items are self-documenting. The third item can be explained as follows.
Our database of the experiment and all results is contained in a unix directory tree
rooted at {\tt exp}.

Inside directory {\tt exp}  
one finds further directories, as indicated below:
\begin{verbatim}
Nuc_CVX_20120605a
Nuc_CVX_20120607a
...
Nuc_CVX_20121107a
Nuc_CVX_20121113a
...
Nuc_CVX_20121120a
Nuc_CVX_20121121a
...
SDP_CVX_20121230h
SDP_CVX_20121230i
..
Nuc_CVX_20130110a
Nuc_CVX_20130110c
...
Nuc_CVX_20130117Cc
Nuc_CVX_20130117Cd
...
Nuc_CVX_20130126Df
Nuc_CVX_20130126Dg
Nuc_CVX_20130126Dh
\end{verbatim}
These directory names are precisely the {\tt Project} names used in the data deposition.
Say we look inside one of these directories, for example 'Nuc\_CVX\_20121121b'. We will
find a directory called {\tt bin} containing software, and a list
of further directories. We excerpt from a  2-column listing of those
directories:

\begin{verbatim}
Nuc_CVX_N05S1a			Nuc_CVX_N30S1c
Nuc_CVX_N05S1b			Nuc_CVX_N30S1d
Nuc_CVX_N05S1c			Nuc_CVX_N30S1e
...
Nuc_CVX_N10S1j			Nuc_CVX_N35S1l
Nuc_CVX_N10S1k			Nuc_CVX_N35S1m
Nuc_CVX_N10S1l			Nuc_CVX_N35S1n
...
Nuc_CVX_N15S1h			Nuc_CVX_N40S1j
Nuc_CVX_N15S1i			Nuc_CVX_N40S1k
Nuc_CVX_N15S1j			Nuc_CVX_N40S1l
...
Nuc_CVX_N20S1o			Nuc_CVX_N45S1q
Nuc_CVX_N20S1p			Nuc_CVX_N45S1r
Nuc_CVX_N20S1q			Nuc_CVX_N45S1s
...
Nuc_CVX_N25S1n			Nuc_CVX_N50S1p
Nuc_CVX_N25S1o			Nuc_CVX_N50S1q
Nuc_CVX_N25S1p			Nuc_CVX_N50S1r
...
\end{verbatim}
These directory names are precisely the {\tt 'Experiment'} values seen in the data deposition.
Inside the {\tt bin} directory we find the matlab code used in common by all the above experiments.
\begin{verbatim}
Nuc_CVX_20121121b$ ls bin
aveRank.m			solveNuc_CVX_Stack_Arb.m
predNucPT.m			stackRankRMatrix.m
rankRMatrix.m
\end{verbatim}
Inside one of the experiment directories we find experiment-specific files (in
both Matlab and Bash script)
as well as output files. The subdirectory {\tt logs} contains logs that were
created while the jobs were running.
\begin{verbatim}
Nuc_CVX_20121121b $ ls -R -C1 Nuc_CVX_N25S1o
bashMain.sh
PTExperiment.m	
matlabMain.m	
results.mat
randState.mat	

Nuc_CVX_N25S1o/logs:
Nuc_CVX_N25S1o.stderr			
Nuc_CVX_N25S1o.stdout
runMatlab.20121121b_Nuc_CVX_N25S1o.log
\end{verbatim}
Here the {\tt .mat} files were produced by matlab during the running
of the experiment.
\bitem
 \item {\tt randState.mat} preserves the state of the random number generator at the beginning of that experiment.
 \item {\tt results.mat} gives the results of the individual problem instances; typically in the form
    $(r,n,N,M,Err0,Err1,Err2)$.
\eitem

\end{document}